\documentstyle[11pt]{article}

\def\be{\begin{equation}}
\def\ee{\end{equation}}
\def\bea{\begin{eqnarray}}
\def\eea{\end{eqnarray}}

\begin{document}

\section*{\center SEARCH FOR THE NEUTRINO MAGNETIC MOMENT IN 
THE NON-EQUILIBRIUM REACTOR ANTINEUTRINO ENERGY SPECTRUM}

\Large
{\hspace{1.5 cm} V.I. Kopeikin, L.A. Mikaelyan, V.V. Sinev}
\\

\normalsize
{\it \center RRC "Kurchatov Institute", Kurchatov Sq., 1, Moscow-123182, Russia.\\
  Resume of seminar talks given at Kurchatov Institute, March 1999 y.}
\hspace{3 cm} 
\vspace{1em}\\
\righthyphenmin=3
 We study the time evolution of the typical nuclear reactor 
antineutrino energy spectrum during reactor ON period and the 
decay of the residual antineutrino spectrum after reactor is stopped.
We find that relevant varia-tions of the soft recoil electron spectra 
produced via weak and magnetic ${\widetilde {\nu}}_{e},e$ scattering 
process can play a significant role in the current and planned searches for the neutrino 
magnetic moment at reactors.

\section{Introduction}
\large
Efforts are currently being done to observe the neutrino magnetic moment below the limit 
${\mu}_{\nu}<2\cdot10^{-10}{\mu}_{B}$ that was found in previous 
${\widetilde {\nu}}_{e},e$ scattering experiments at SAVANNAH RIVER, KRASNOYARSK and 
ROVNO reactors [1]. The KURCHATOV-PNPI collaboration 
is plan-ning for KRASNOYARSK new studies of low kinetic energy recoil electrons 
in ${\widetilde {\nu}}_{e},e$ experiment with a Si semiconductor multi detector. 
The MUNU collaboration experiment at BUGEY with a gas TPC chamber is in
the final state of preparation [2]. 

The dominant contribution to the soft recoil electron produced 
in the ${\widetilde {\nu}}_{e},e$ scattering comes from the low energy part of 
the reactor ${\widetilde {\nu}}_{e}$ energy spectrum. This part of the spectrum 
is strongly time dependent: it never comes to saturation during the reactor operating 
run and does not vanish after the reactor is shut down, the time when 
the background is usually measured.

Here we consider the time evolution of the typical reactor ${\widetilde {\nu}}_{e}$
energy spectrum and discuss relevant variations of the ${\widetilde {\nu}}_{e},e$ 
scattering recoil electron spectra.

\section{TIME VARIATION OF THE REACTOR \\ANTINEUTRINO SPECTRUM}

1. Three components contribute the reactor ${\widetilde {\nu}}_{e}$ energy spectrum \\
${\rho}(E)/fiss\cdot MeV$: 
\begin{equation}
{\rho}(E)=^{F}{\rho}(E) + ^{U}{\rho}(E) + {\Delta}{\rho}(E).
\end{equation}
Here, the term $^{F}{\rho}(E)$ represents the radiation of the $^{235}U$, $^{239}Pu$,
$^{238}U$ and $^{241}Pu$ fission fragments. The second term stems from the 
chain of the ${\beta}$-decays which follow neutron radiative capture in $^{238}U$:
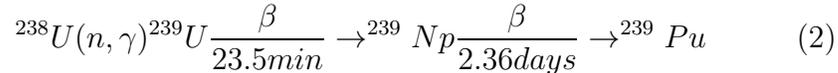
\begin{equation}
^{238}U(n,{\gamma})^{239}U \frac{\beta}{23.5 min}\to ^{239}Np \frac{\beta}{2.36 days}\to ^{239}Pu
\end{equation}
The last term in Eq.(1) accounts for the antineutrinos (and neutrinos) induced by 
the neutron interactions with other materials in the reactor core. 
As discussed in Ref.[3] this term adds no more than 1\% to 
the total reactor ${\widetilde {\nu}}_{e}$ flux and is disregarded here. 

Till recently the term $^{F}{\rho}$ has traditionally been identified with the 
reactor ${\widetilde {\nu}}_{e}$ spectrum. The contribution of the chain (2) 
antineutrinos is however quite sizable for all reactors where neutrino experiments are 
running or planned. In the ROVNO, BUGEY and CHOOZ PWR-type reactors 
about 1.2 ${\widetilde {\nu}}_{e}$ per fission come from this source.
\vspace{0.2 cm}

2. For each of the four isotopes $^{235}U$, $^{239}Pu$, $^{238}U$ and $^{241}Pu$ the 
evolution of the neutrino spectra ${\rho}(E,t)$ have been calculated vs time t since 
the beginning of the fission process. The subsequent decay of the spectra during 
reactor OFF period have been followed. 

The base used for these calculations involves data on 571 fission fragments, data on 
nuclear isomers and delayed neutron emission.
\vspace{0.2 cm}

3. Calculations show that in PWR reactors about 2/3 of all antineu-trinos belong 
typically to the energy range below E = 1,5 MeV. This part of the 
${\widetilde {\nu}}_{e},e$ spectrum ${\rho}(E,330)$ and it's component due to fission 
fragments ${\rho}(E,330)$ at the end of the reactor 330 day ON period are 
presented in Fig.1.
 
The evolution of the ${\widetilde {\nu}}_{e},e$ spectrum  during PWR reactor ON period 
and its decay after the reactor is shut down is illustrated in Fig.2a,b. 

\section{RECOIL ELECTRON ENERGY SPECTRA}

1. The recoil-electron spectra $S^{W}$(T) and $S^{M}$(T) in $cm^{2}/MeV\cdot fiss.$
units, (T is the recoil-electron kinetic energy) for weak (W) and magnetic (M) 
scattering of reactor antineutrinos are found by convolution the ${\widetilde {\nu}}_{e},e$ 
spectra ${\rho}(E)$ with the differential cross sections for monoenergy antineu-trino:
\begin{equation}
\frac{d{\sigma}^{W}}{dT}=g^{2}_{F}\frac{m}{2{\pi}}\cdot [4x^{4}+(1+2x^{2})(1-\frac{T}{E})^{2}-2x^{2}(1+2x^{2})\frac{mT}{E^{2}}]
\end{equation}
\begin{equation}
\frac{d{\sigma}^{M}}{dT}={\pi}r^{2}_{0}\frac{{\mu}^{2}_{\nu}}{{\mu}^{2}_{B}}(\frac{1}{T}-\frac{1}{E}),
\end{equation}
where m is the electron mass, $g^{2}_{F}\frac{m}{2{\pi}}=4.31\cdot 10^{-45} cm^{2}/MeV$, 
$x=sin^{2}{\theta}_{W}=0.232$ is the Weinberg parameter, 
${\pi}r^{2}_{0}=2.495\cdot 10^{-25} cm^{2}$.
\vspace{0.2 cm}

2. Calculated recoil-electron spectra $S^{W}$(T,330) and $S^{M}$(T,330) 
at the end of the reactor ON period are shown in Fig.3. In searches
for the neutrino magnetic moment, weak ${\widetilde {\nu}}_{e},e$ scattering plays the 
role of the reactor-correlated background. We note, that in order to keep this 
background at sufficiently low level one should try to study recoil electrons at 
not too high energies. As an example, the recoil electron energies $T > 100$ keV seem 
to be "high" to search for ${\mu}_{\nu} = 2\cdot 10^{-11}{\mu}_{B}$ while the 
range $T < 700$ keV is tolerably low for ${\mu}_{\nu} = 5\cdot 10^{-11}{\mu}_{B}$.

Calculated time variations of the recoil-electron spectra during PWR reactor ON 
and OFF periods for weak and magnetic scattering are presented in Fig.4a,b.

\section{DISCUSSION AND CONCLUSIONS}

In practice the situation is not as simple as presented above. There occur deviations
from the standard operating schedule: reactor can be stopped for a few days, or it
can operate at a reduced level of power etc. For each particular experiment a comprehensive 
analysis of the reactor operation data should be carried out including all details.

The main result of this study is that effects due to neutrino relaxa-tion play not
negligible role in sensitive searches for the neutrino magne-tic moment at reactors.

\section*{Acknowledgments}
  
This work is supported by RFBR, projects 97-02-16031, 96-15-96640.


\end{document}